\documentclass[conference]{./IEEEtran}

\usepackage{url}            
\usepackage{booktabs}       
\usepackage{amsfonts}       
\usepackage{nicefrac}       
\usepackage{microtype}      
\usepackage{blindtext, graphicx}
\usepackage[cmex10]{amsmath}
\usepackage{amssymb}
\usepackage{mathtools}
\usepackage{cite}
\usepackage{todonotes}
\usepackage{pgfplots}
\pgfplotsset{compat=newest}
\usepackage{pgfplotstable}
\usepackage[caption=false,font=footnotesize]{subfig}

\graphicspath{{figures/}}
\DeclareGraphicsExtensions{.pdf}

\IEEEoverridecommandlockouts
\begin{document}
\newcommand{\OurScheme}{LogicNets}
\title{\OurScheme{}: Co-Designed Neural Networks and Circuits for
Extreme-Throughput Applications}

\author{\IEEEauthorblockN{Yaman Umuroglu\footnotemark{*}\thanks{\footnotemark{*}Equal contribution}, Yash Akhauri\footnotemark{*}, Nicholas J. Fraser and
Michaela Blott}
\IEEEauthorblockA{Xilinx Research Labs\\
Dublin, Ireland
}}

\newcommand{\TODO}[1]{\todo[inline]{#1}}

\maketitle

\begin{abstract}
Deployment of deep neural networks for applications that require very high
throughput or extremely low latency is a severe computational challenge,
further exacerbated by inefficiencies in mapping the computation to hardware.
We present a novel method for designing neural network topologies that directly
map to a highly efficient FPGA implementation.
By exploiting the equivalence of artificial neurons with quantized
inputs/outputs and truth tables, we can train quantized neural networks that can
be directly converted to a netlist of truth tables, and subsequently deployed as
a highly pipelinable, massively parallel FPGA circuit.
However, the neural network topology requires careful consideration since
the hardware cost of truth tables grows exponentially with neuron fan-in.
To obtain smaller networks where the whole netlist can be placed-and-routed
onto a single FPGA, we derive a fan-in driven hardware cost model to guide
topology design, and combine high sparsity with low-bit activation quantization
to limit the neuron fan-in.
We evaluate our approach on two tasks with very high intrinsic throughput
requirements in high-energy physics and network intrusion detection.
We show that the combination of sparsity and low-bit activation quantization
results in high-speed circuits with small logic depth and low LUT cost,
demonstrating competitive accuracy with less than 15~ns of inference latency and
throughput in the hundreds of millions of inferences per second.
\end{abstract}

\section{Introduction}
\label{sec:intro}

Deep Neural Networks (DNNs) have a wide application scope beyond the highly-popular computer vision use cases, promising to replace manual algorithmic
implementations in many application domains.
Certain applications, such as data collection from particle physics
experiments \cite{duarte2018fast}, line-rate filtering of packets for network
intrusion detection~\cite{murovivc2019massively} and wireless
communications~\cite{Shi2019DeepLF}, have stringent real-time requirements and
extremely high data-rates.
Replacing parts of such \emph{extreme-throughput} applications with machine-learned components requires highly specialized DNN implementations that offer
inference rates reaching hundreds of millions of samples per second with sub-microsecond latency.
This is the key challenge we try to address in this work: how can we build DNN
implementations that are able to meet the performance and latency constraints
for extreme-throughput applications?

Popular hardware platform options for accelerating DNN inference include GPGPUs,
FPGA overlays and specialized tensor processors \cite{blott2019qutibench}.
Most of these alternatives apply the traditional paradigm in computer design
where hardware and software are designed separately, bridged by a compiler
that generates instructions to schedule the required computation onto available
hardware.
However, this flexibility typically comes at the cost of performance overheads,
making it difficult to apply this approach to extreme-throughput applications.
Prior work \cite{umuroglu+:FPGA2017finn, tridgell2019unrolling, wang2019lutnet, duarte2018fast}
demonstrated that specialized co-design approaches are able to produce
FPGA DNN implementations that yield increased throughput, while still offering the ability
to reconfigure to address changing requirements.

In this paper, we present a novel method named \OurScheme{} for co-designing DNN
topologies that map directly to an efficient FPGA implementation for
extreme-throughput applications.
Our scheme is based on the observation that artificial neurons with quantized
inputs and outputs can be converted to truth tables.
However, an efficient FPGA implementation for a truth table is generally only
possible when the number of inputs is small.
By limiting neuron fan-in using activation quantization and sparsity, we show
how we can design DNN topologies that are still trainable using
standard backpropagation, and can be mapped directly to an equivalent
hardware circuit with small combinatorial depth.
DNNs designed and trained in this manner result in fast and efficient FPGA implementations that can fulfill the performance
requirements for extreme-throughput applications.
Extending on our abstract in~\cite{umuroglu2020logicnets_fccm}, this paper
makes the following contributions:
\begin{itemize}
    \item We describe \OurScheme{}, a DNN-hardware co-design methodology
    that allows trained quantized networks to be directly converted to an equivalent hardware netlist of truth tables.
    \item We exploit sparsity and quantization to reduce the neuron fan-in, and
    provide an analytical cost model to quickly estimate the required FPGA resources to guide topology design.
    \item We develop a PyTorch library to train sparse, quantized topologies and
    convert them to Verilog netlists.
    \item We empirically evaluate our approach on two extreme-throughput tasks, demonstrating FPGA implementations with competitive accuracy and
    throughput in the hundreds of millions of inferences per second.
\end{itemize}


\section{Background}
\label{sec:bkg}

In this section, we briefly describe
prior work relating to DNN inference acceleration on FPGAs,
and schemes previously used to construct quantized and sparsely-connected DNNs.

\subsection{Sparse and Quantized Neural Networks}
\label{sec:sparsednn}
In a sparse neural network, each layer of neurons receives inputs from only a
few connections of the previous layer of activations, in contrast to dense
networks where all previous layer activations are inputs to each neuron of the
next layer.
Numerous techniques to build sparse DNN topologies exist, including learned
sparsity \cite{Wortsman2019DiscoveringNW}, pruning techniques \cite{han2015deep}
and a priori fixed sparsity \cite{prabhu2018deep} which we utilize in this work
due to its relative simplicity.
Quantization involves restricting weights, activations or both to a set of
discrete values. To preserve DNN accuracy with low-bit quantization (e.g. $\leq 4\mathrm{-bits}$) it is typically necessary
to use specialized techniques during training, such as using the Straight-Through Estimator (STE)~\cite{bengio2013estimating}
to propagate gradients through non-differentiable quantizers and learned scale factors to reduce approximation error.
We refer the reader to the survey by Guo~et~al.~\cite{DBLP:journals/corr/abs-1808-04752} for further details.

\subsection{Prior Work on FPGA DNN Inference}

\begin{figure}[!t]
		\centering
		\subfloat[LUTNet-style~\cite{wang2019lutnet, murovivc2019massively} with weights in LUT equations. An explicit accumulation and activation datapath is still present.]{\includegraphics[width=0.45\textwidth]{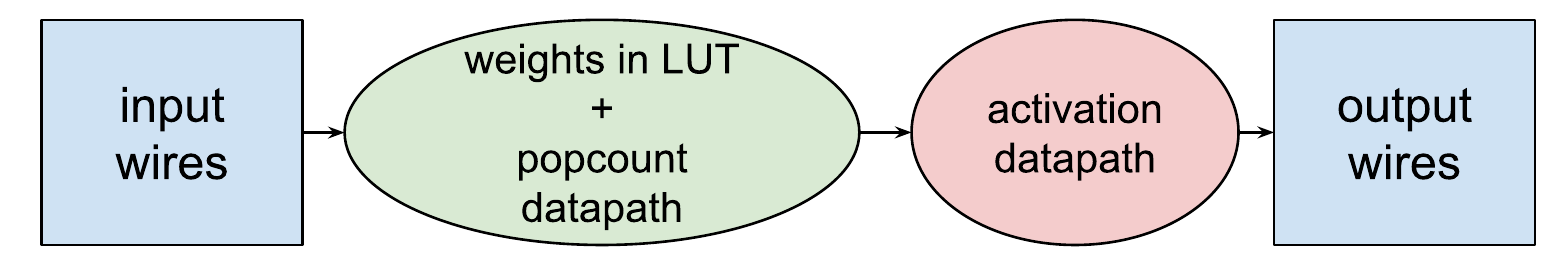}
		\label{fig:accel-lutnet}}
		\hfil
		\subfloat[\OurScheme{} and NullaNet~\cite{nazemi2018nullanet} with all  operations packed into LUTs.]{\includegraphics[width=0.45\textwidth]{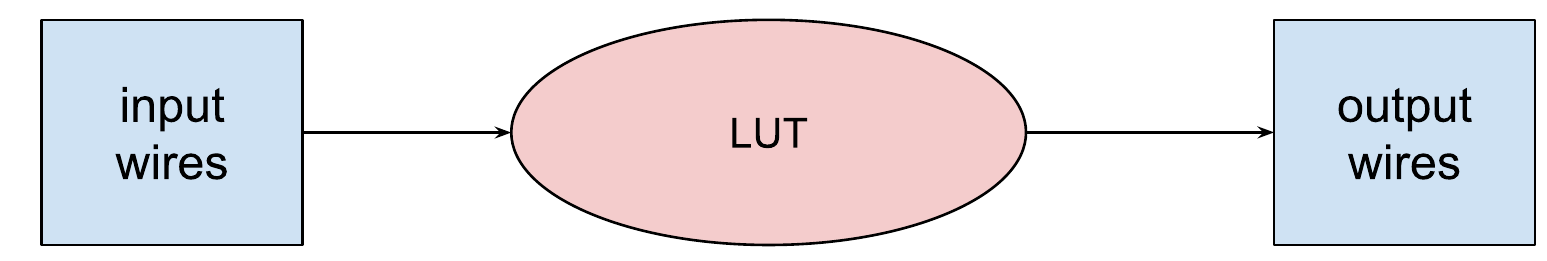}
		\label{fig:accel-fpganet}}
		\caption{Two architectural alternatives for specialized FPGA inference.}
		\label{fig:arch-alternatives}
\end{figure}

DNN inference typically consists of applying a sequence of multiply-accumulates
(MACs),
followed by a nonlinear activation function.
Many alternatives exist when mapping these computations to the FPGA fabric and a large body of prior work exists on FPGA DNN
inference; here, we only cover the works closely related to ours
and refer the reader to a recent survey by Zhao~et~al.\cite{zhao2018hardware} for further reading.
We organize our discussion of prior work according to the presence or absence of explicit weight storage, MAC and activation datapaths
in the proposed architecture, as illustrated in Figure~\ref{fig:arch-alternatives}.


\emph{Weights in LUT equations.}
Figure~\ref{fig:accel-lutnet} shows architectures with network weights ``baked in'' into LUT equations, but parts of the MAC and activation datapaths still exist.
This enables greater performance with fully unrolled (non-time-multiplexed) implementations without a control path, but is less flexible.
Wang~et~al.~\cite{wang2019lutnet} introduce LUTNet, a LUT-optimized FPGA inference scheme which achieves high LUT density.
\cite{wang2019lutnet} take a pruned version of ReBNet~\cite{ghasemzadeh2018rebnet} in which some of the XNOR-popcount operations are mapped more effectively to $k$-input LUTs, although explicit popcount and thresholding datapaths are still present and occupy significant resources.
Murovic~et~al.~\cite{murovivc2019massively} implement binarized networks which have been fully unrolled and implemented directly into LUTs of a small FPGA,
although explicit accumulation and activation datapaths are still present.
This unrolling allow the synthesis optimization tool to potentially simplify significant portions of the compute logic.
Duarte~et~al.~\cite{duarte2018fast} present a package called hls4ml which generates high-level synthesis-based FPGA designs which supports two axes of folding, but can also generate full-unrolled designs.
\cite{duarte2018fast} operates at higher bit-widths (8 or 16-bit) and maps significant portions of the compute to the DSP blocks.

\emph{No explicit datapath.}
Figure~\ref{fig:accel-fpganet} shows the architecture proposed in this work, where all operations and weights are packed into a truth table
and no explicit MAC or activation datapath is present.
Nazemi~et~al.~\cite{nazemi2018nullanet} introduce NullaNet, which proposes converting activation-quantized neural networks into large truth tables in a similar fashion to this work.
Their stated goal is reducing the number of memory accesses, whereas \OurScheme{} aims to co-design DNNs that can yield FPGA circuits that offer extremely high throughput and low latency.
Key differences between NullaNet and this work are as follows:
\begin{itemize}
\item NullaNet only considers densely connected networks and suffers from high fan-in, we use sparse topologies to avoid this problem (Section~\ref{sec:restrict-fanin}).
\item NullaNet uses a lossy truth table sampling method to overcome the fan-in problem which gives an approximation of the DNN, whereas we use a lossless method (Section~\ref{sec:neq-hbb-intro}).
\item NullaNet only considers binary quantization of activations, we consider several low-precision variants, which is key to achieving higher accuracy (Section~\ref{sec:act_bw_impact}).
\end{itemize}

\section{Building DNNs from Small Truth Tables}
\label{sec:method-theory}
In this section, we explain the core concepts that form the foundations for
\OurScheme{}.
As the performance demands for extreme-throughput applications are so high,
the available hardware must be utilized to its full potential.
We aim to answer the following question: \emph{given that the building blocks
of FPGA hardware are small truth tables that can implement \textbf{any} function,
how can we design DNN topologies that map well to these building blocks?}

\subsection{Equivalence of Neurons and Truth Tables}
\label{sec:neq-hbb-intro}

\begin{figure}[!t]
		\centering
    \includegraphics[height=2.5cm]{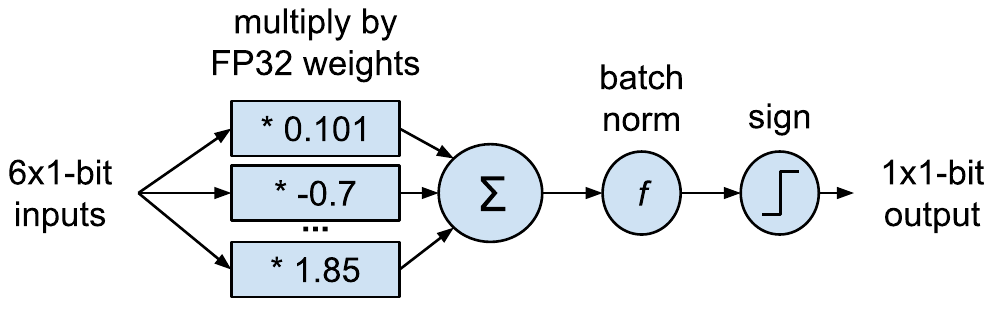}
		\caption{A 6:1 NEQ with batch normalization that maps to a 6:1 LUT.}
		\label{fig:neqs}
\end{figure}

The foundation of \OurScheme{} is the equivalence of artifical neurons with
quantized inputs/outputs and truth tables.
Consider an artifical neuron with $C_{in}$ different inputs, where each input
is $\beta$ bits wide, and let the neuron produce a single $\beta$-bit output.
Let $X$ be the total number of input bits or \emph{fan-in}, \emph{i.e.} $X = \sum_{i=0}^{C_{in}}
\beta$ and $Y=\beta$ be the total number of output bits,
Regardless of the internal complexity of the neuron, we can always implement
its functionality with an $X$-input, $Y$-output (denoted $X:Y$) truth table
by enumerating all of its possible $2^X$ different inputs, observing
the output, and recording it into the truth table.

In \OurScheme{}, we refer to the Verilog implementation of an $X:Y$ truth
table as a \emph{Hardware Building Block (HBB)}, and any trained artifical neuron
that can be converted into an HBB as a \emph{Neuron Equivalent (NEQ)}.
Since the internal complexity of NEQs does not matter as long as the number of
input-output bits are fixed, we can add components that makes the DNN training
process easier into the NEQ.
Figure~\ref{fig:neqs} illustrates this for a 6:1 NEQ with floating point
weights and batch normalization, with six FP32 values for the weights
and four for the batch normalization parameters, requiring $10 \cdot 32 = 320$
bits of storage.
The equivalent HBB only requires 64 bits to store the truth table, which is
$5\times$ smaller.

\subsection{The FPGA LUT Cost of Large Truth Tables }
To implement NEQs that have either more inputs or more bits per input, we
need HBBs with larger $X$ values, which can be implemented by combinations of
smaller FPGA LUTs.
For instance, the output of two 6:1 LUTs can be fed into a third 6:1 LUT to
act as a multiplexer, implementing a 7:1 LUTs using three 6:1 LUTs.
For higher output bitwidth $Y$ the cost scales linearly, if each additional
output bit is produced by adding a copy of the table.
Generalizing this, we can analytically estimate the number of 6:1 FPGA LUTs
required to implement an $X:Y$ HBB using
Equation~\ref{eqn:lutcost}.
\begin{equation}
	\label{eqn:lutcost}
	\mathrm{LUTCost(X, Y)} = \frac{Y}{3} \cdot (2^{X-4} - (-1)^{X})
\end{equation}
However, scaling up $X$ in this manner is expensive: the number of LUTs
needed grows exponentially\footnote{Although special FPGA
capabilities such as F7MUX/F8MUX and heuristic logic minimization decrease the
LUT cost, the exponential trend remains.} with $X$.
For instance, implementing a single $32:1$ NEQ would require close to a hundred
million 6:1 LUTs, which is much larger than even the largest FPGAs available
today.
Thus, it is critical to keep the fan-in $X$ small enough so that each NEQ in the
topology has a reasonably small LUT cost.

\subsection{Designing Topologies with Restricted Fan-In}
\label{sec:restrict-fanin}
\begin{figure}
		\centering
		\includegraphics[width=0.45\textwidth]{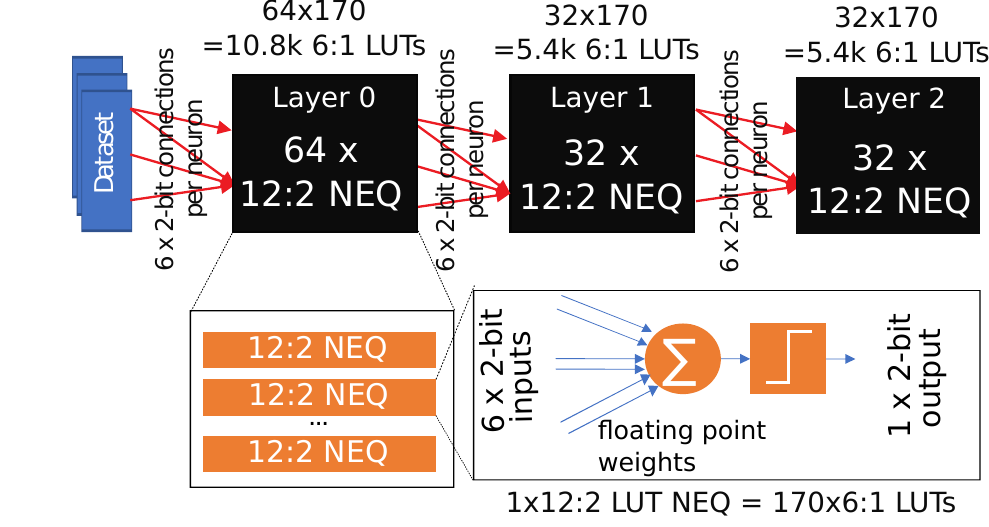}
		\caption{Example sparse network and 6:1 LUT cost calculation.}
		\label{fig:hw-cost-example}
\end{figure}

Modern DNN topologies do not necessarily restrict neuron fan-in, which makes
them impractical for direct mapping to HBBs.
In \OurScheme{}, we propose to \emph{co-design} DNN topologies with awareness
of the FPGA hardware implementation cost by restricting the fan-in of each
artifical neuron.
Recalling that the fan-in is computed as $X=\sum_{i=0}^{C_{in}} \beta$, we
observe that we have two key topological parameters to control it:

\begin{itemize}
  \item \textbf{Number of inputs $C_{in}$:} A DNN may have hundreds of
  neurons per layer, and connecting a neuron's inputs to every other neuron
  in the previous layer will result in an intractibly large fan-in.
  We take inspiration from the developments in sparse
  topologies (Section~\ref{sec:sparsednn}), connecting each neuron to $\gamma$ previous neurons, with a
  $\gamma$ much smaller than the previous layer size.

  \item \textbf{Bitwidth of inputs $\beta$:} The bitwidth of the activations
  from the previous layer also has a large impact on fan-in $X$.
  We apply training-time techniques developed in prior work (Section \ref{sec:sparsednn}) to quantize the activations to $\leq 4$-bits, which reduces the fan-in substantially.
\end{itemize}

With these fan-in restrictions in place, we can explore topologies with
different number of neurons and layers while avoiding the exponential
growth in LUT cost, using the cost model to guide the exploration.
To compute the total FPGA resource cost for a network, we can simply sum
the resources taken up by each NEQ in the topology.
We provide an example for the network illustrated in
Figure~\ref{fig:hw-cost-example}.
Here, each neuron output is quantized to $\beta=2$ bits, and is connected to
$\gamma=6$ outputs
from the previous layer as its input.
Thus, $X = 6 \cdot 2 = 12$ and $Y = 2$.
According to Equation~\ref{eqn:lutcost}, the cost for a 12-bit-input,
2-bit-output NEQ is
$\mathrm{LUTCost(12, 2)}
= \frac{2}{3} \cdot (2^{12-4} - (-1)^{12})
= 170$~6:1~LUTs.
Since there are 128 such NEQs in total, the estimated cost for this network is
$128 \cdot 170 = 21760$~6:1~LUTs.

\section{The \OurScheme{} Design Flow and Implementation}
\label{sec:method}

Having explored the foundational ideas in Section~\ref{sec:method-theory},
we now present \OurScheme{} as a three-step design flow to train DNNs that map
directly and efficiently to FPGAs:

\begin{enumerate}
\item Define Hardware Building Blocks (HBB) and Neuron Equivalents (NEQ)
\item Define and train a DNN of NEQs in PyTorch then convert to netlist of HBBs
\item Postprocess the netlist, synthesize to obtain a bitfile
\end{enumerate}

\begin{figure}[!t]
		\centering
		\subfloat[Define HBBs and NEQs]{\includegraphics[width=0.45\textwidth]{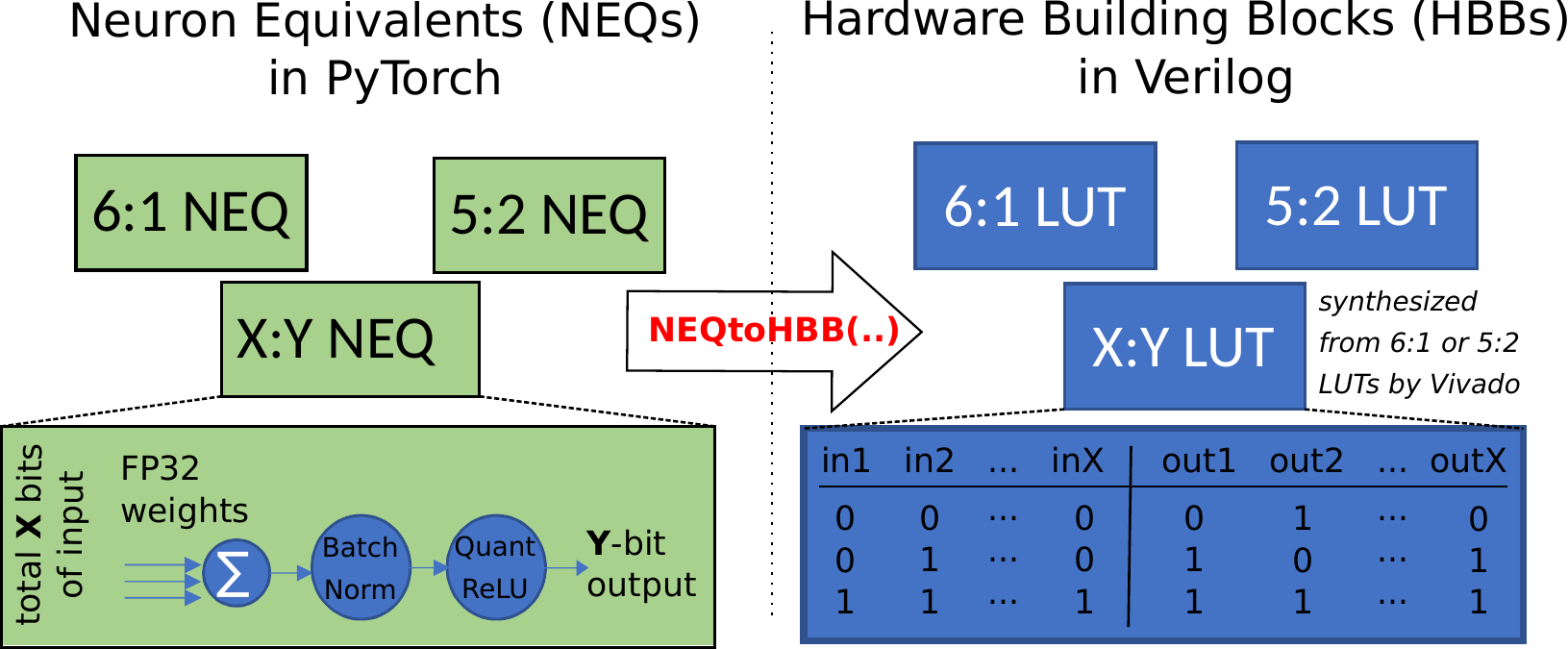}
		\label{fig:hbbneq}}
		\hfil
		\subfloat[Train and convert a network of NEQs]{\includegraphics[width=0.45\textwidth]{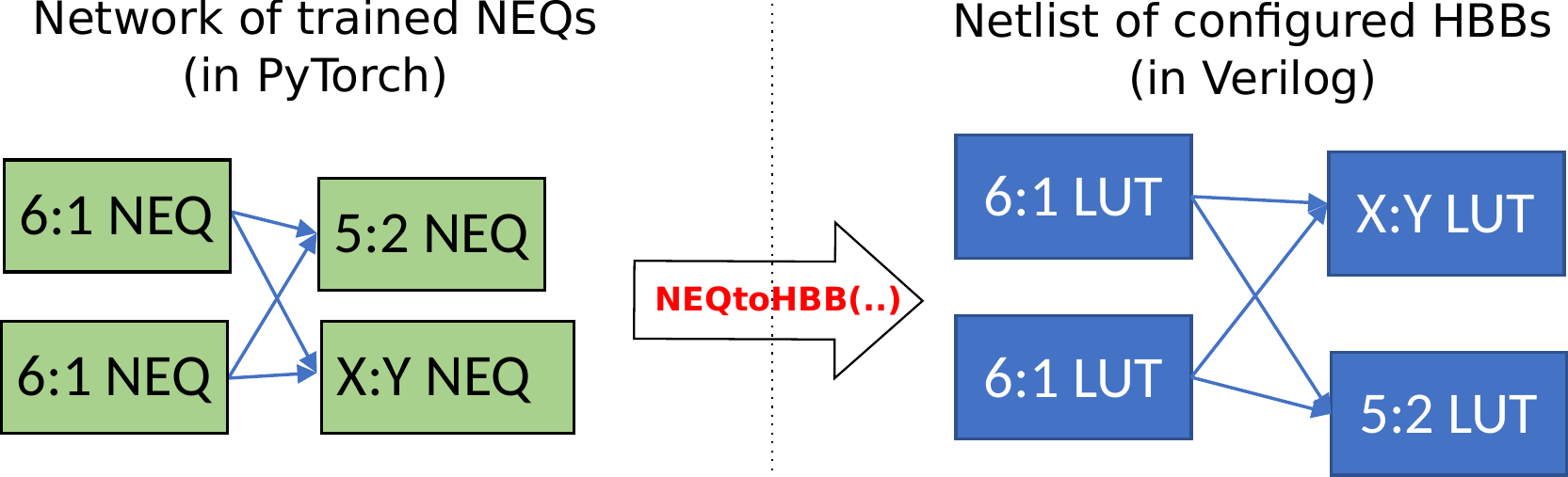}
		\label{fig:neqtohbb}}
		\hfil
		\subfloat[Postprocessing and deployment]{\includegraphics[width=0.45\textwidth]{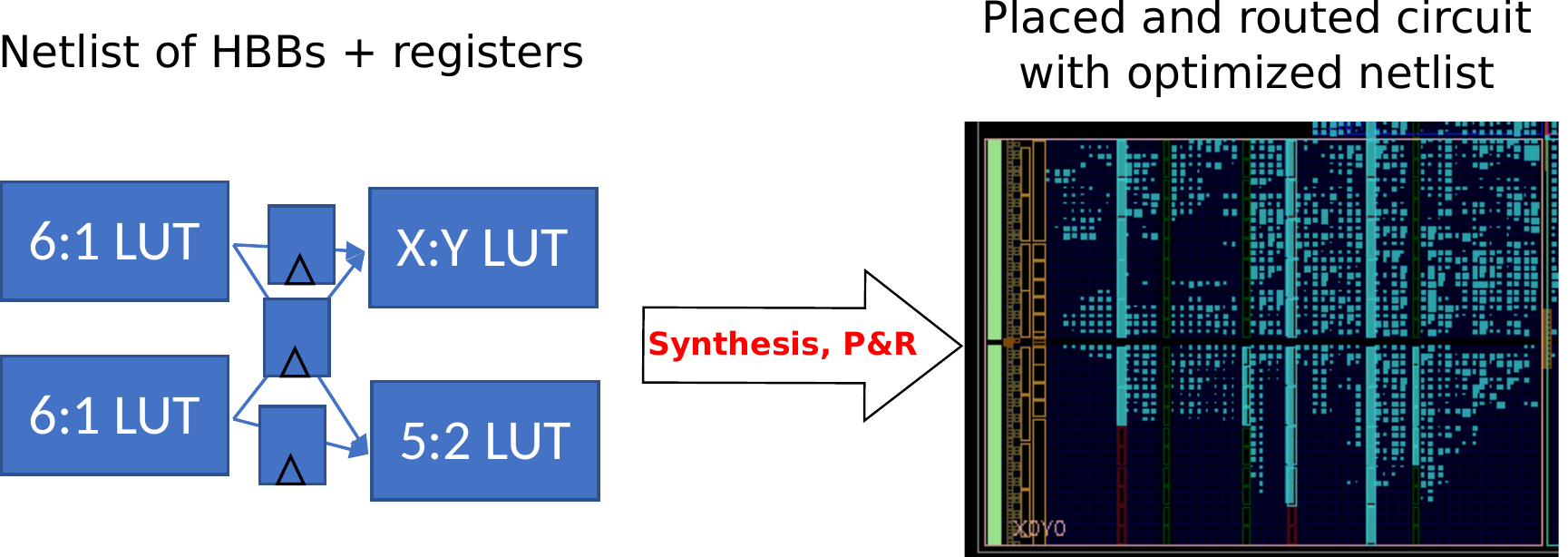}
		\label{fig:postprocdeploy}}
		\caption{Three steps in the \OurScheme{} design flow.}
		\label{fig:schemeoverview}
\end{figure}

Figure~\ref{fig:schemeoverview} illustrates the steps in the \OurScheme{} design
flow.
We have implemented a prototype of this design flow with a PyTorch library in
order to enable faster design space exploration.
In the following sections, we describe each step in greater detail.

\subsection{Define HBBs and NEQs}

As described in Section~\ref{sec:neq-hbb-intro}, NEQs and HBBs constitute
the building blocks of \OurScheme{} on the PyTorch and Verilog side,
respectively.
The first step in the design flow is to identify the range of $X:Y$ values
that yield HBBs with reasonable LUT cost, and to define corresponding NEQs
in PyTorch that can be trained with the sparsity and activation quantization
restrictions in place.
These can be NEQs that map to single 6:1 or 5:2 FPGA LUTs, or a generic X:Y
truth table that will be implemented by the synthesis tool.
In this work, we add batch normalization followed by uniform quantization with
learned scale factors using Brevitas~\cite{xilinx2019brevitas} for better
accuracy and easier training.
The first step needs to be performed only once per FPGA device family,
since the identified building blocks can be used to construct multiple
different topologies.

\subsection{Train and convert a network of NEQs}
\label{sec:training}
The second step is followed for each new DNN that needs to be trained and
deployed, and takes place in the PyTorch machine learning framework.
Using the available NEQs identified in Step 1, a deep neural network topology is
constructed by instantiating NEQs and connecting them together.
We use the approach described in \cite{prabhu2018deep} to provide
fixed random sparsity with the desired per-neuron fan-in.
To guide topology design, our library implements the analytical model from
Equation~\ref{eqn:lutcost} to estimate the required FPGA resources prior to training.
Once the topology is defined, the DNN is trained in PyTorch using standard
DNN optimizers and backpropagation.
Methods applied to improve standard DNN training such as knowledge
distillation and ensembling can also be applied here.
Finally, the trained network is converted into a Verilog netlist
of HBB instances and their (sparse) connections.
To convert NEQs into HBBs, we follow the enumeration-based procedure
in Section~\ref{sec:neq-hbb-intro} to evaluate each input
combination and add an entry into the HBB truth table.
The truth tables are expressed as read-only memories (ROMs) with a \texttt{case}
statement returning the evaluated constant for each input combination,
and we leave it to the synthesis tool to map the ROMs to FPGA LUTs.

\subsection{Postprocessing, Synthesis and Deployment}

Any optimization admitted by a netlist can be applied as the postprocessing
step.
For instance, a heuristic logic minimizer can be applied to the network to use
fewer LUTs, pipeline registers can be inserted between the layers to increase
the clock frequency, or the netlist can be split up into chunks for mapping to
a smaller FPGA with dynamic partial reconfiguration, one chunk at a time.
In this paper, we focus on single-FPGA implementations for extreme-throughput
applications and only consider register insertion between layers for
postprocessing, as shown in Figure~\ref{fig:postprocdeploy}.
After any preprocessing is complete, the final netlist is processed
with synthesis, place-and-route algorithms to yield an FPGA bitfile.
As our results in Section~\ref{sec:lut_reduction_synth_opts} indicate, synthesis-time
optimizations such as heuristic logic minimization can yield significant
hardware cost savings, effectively pruning the network further to make it more
sparse.

\section{Evaluation}

\begin{table*}[t]
\begin{center}
    \caption{Highlights from \OurScheme{} results on the chosen tasks.}
    \label{tab:key_results}
    \begin{tabular}{lllllrrrrl}
    \toprule
    Name & Neurons per Layer             & $\beta$ & $\gamma$ & Accuracy & Model LUT & Synth LUT    & FF    & Reported $F_{\mathrm{max}}$ & Remarks                            \\
    \midrule
    JSC-S & 64, 32, 32, 32       & 2       & 3        & 67.8\%  & 330     & 214    & 244   & 1,585~MHz           &                                    \\
    JSC-M & 64, 32, 32, 32       & 3       & 4        & 70.6\%  & 42,075  & 14,428 & 440   & 599~MHz             &                                    \\
    JSC-L & 32, 64, 192, 192, 16 & 3       & 4        & 71.8\%   & 303,285 & 37,931 & 810   & 427~MHz             & $\beta_i=4, \beta_o=7, \gamma_o=5$ \\
    \midrule
    NID-S & 593, 100             & 2       & 7        & 83.88\%    & 473,308   & 3,586  & 1,320 & 811~MHz             &                                    \\
    NID-M & 593, 256, 128, 128   & 2       & 7        & 91.30\%    & 754,292   & 15,949 & 1,274 & 471~MHz             &                                    \\
    NID-L & 593, 100, 100, 100   & 3       & 5        & 88.68\%    & 1,021,175 & 25,050 & 1,421 & 418~MHz             & $\beta_i=2, \gamma_i=7$            \\
    \bottomrule
    \end{tabular}
\end{center}
\end{table*}\textbf{}


In order to evaluate \OurScheme{} we picked the
following tasks with extreme-throughput requirements from two domains:
\begin{itemize}
\item \emph{Jet Substructure Classification (JSC)}: Large-scale physics
experiments such as those in CERN produce terabytes of
instrumentation data every second, which is processed by a
hierarchy of \emph{triggers} to filter out the interesting
results. Recent work by Duarte~et~al.~\cite{duarte2018fast}
successfully applied DNNs as to the Jet Substructure
Classification (JSC) task. This task targets the FPGA-based
triggers of the CERN ATLAS and CMS experiments, which must be
pipelined to handle a data rate of 40 MHz and limit response latency to
less than a microsecond.
We use the formulation from Duarte~et~al.~\cite{duarte2018fast}
for JSC as a 16-input, 5-output classification task,
and refer the reader to their work for a more detailed
explanation of the task.

\item \emph{Network Intrusion Detection (NID)}: FPGAs are commonly used for implementing
high-performance packet processing systems that still provide a degree of
programmability \cite{xilinx2014sdnet}.
An advantage of such systems is their ability to facilitate stronger network
security by detecting malicious or suspicious network packets, which may be
implemented using DNNs \cite{hb2018deep}.
To avoid introducing bottlenecks on the network, the DNN implementation
must be capable of detecting malicious ones at line rate, which can be millions of packets per second, and is expected to increase further as next-generation networking
solutions provide increased throughput.
To assess \OurScheme{} on this domain we use the UNSWNB15
dataset~\cite{moustafa2015unsw} which provides example packets labeled as bad
(0) or normal (1) with 49 generated input features extracted from simulated modern
intrusion attacks.
We follow the approach by Murovic et al.~\cite{murovivc2019massively} in terms
of dataset preprocessing and feature conversion.
\end{itemize}

We trained a number of
sparsely-connected, activation-quantized MLPs on the chosen tasks using the \OurScheme{} PyTorch library.
All layers use the same $\gamma$ and $\beta$, except when
$\gamma_i$, $\gamma_o$, $\beta_i$ and $\beta_o$  are used to specify the first and last layers' fan-in and bitwidth, respectively.
Based on the feedback from our cost model, we limit $X=\beta \cdot \gamma \leq 15$ in our exploration to focus on \OurScheme{} implementable on a single FPGA.
All networks presented here are trained for 1000 epochs,
with a mini-batch size of 1024, using the ADAM optimizer and a step decay
learning rate schedule starting from $0.1$.
Once the training is complete, we generate
Verilog from the trained network as described in Section~\ref{sec:training}.
We use Xilinx Vivado 2018.3 in out-of-context mode with the default settings for
synthesis and \texttt{Flow\_PerfOptimized\_high} for place and route without any
manual placement constraints, targeting the \texttt{xcvu9p-flgb2104-2-i} FPGA part.
We insert registers at the network input, output and between every layer, and
constrain the clock to 1~ns to achieve the highest possible frequency.
The correctness of the generated hardware is verified by performing
post-synthesis simulation and ensuring the same results as the original PyTorch
network are returned.

\subsection{Overview of Key Results}

Table \ref{tab:key_results} presents several networks obtained using \OurScheme{} on the JSC and NID
tasks, picked to illustrate interesting points from our partial design space exploration.
The table names each datapoint by indicating which task the network was trained on, specifies the network
topology and quantization, the test accuracy, and FPGA resources from both the analytical model and
post-synthesis results.
We make the following observations:

\emph{Scalable resource footprint.} \OurScheme{}-style models expose many topological knobs to control
the size of the network, which translates into FPGA resource savings at the cost of some accuracy.
This is reflected in the variety of neurons per layer, $\beta$ and $\gamma$ in Table~\ref{tab:key_results}.
For instance, JSC-S is able to achieve close to 68\% accuracy using only 214 LUTs, while JSC-L offers
4\% points better accuracy at the cost of 37.9k LUTs.

\emph{Simple, high-frequency circuits.} \OurScheme{} yield simple circuits by design, with as little
as a single level of LUTs between registers when neuron fan-in is constrained to be $\beta \cdot \gamma \leq 6$.
This is the case for JSC-S, which has a reported $F_{\mathrm{max}}$ of 1.5~GHz.
Although this frequency cannot be achieved in practice due to limitations on the global clock network
of the FPGA, the positive slack yielded by the the simplicity of \OurScheme{}-style netlists is still
quite significant and would facilitate timing closure when integrating \OurScheme{}-style DNNs into other
high-speed designs.
Larger \OurScheme{}-designs such as NID-L with $\beta \cdot \gamma = 15$ require more levels of logic and
resources and are more challenging to place and route, but are still capable of achieving high clock rates
over 400~MHz.

\emph{Competitive accuracy with sparsity and quantization.} We are able to obtain datapoints which offer accuracy over 70\%
for the JSC task and over 90\% for the NID task, which is competitive with
the accuracy reported by related work (Section~\ref{sec:comparison_to_related_work}).
We expect accuracy to further increase by using better sparsity methods
compared to fixed random sparsity, such as magnitude pruning and sparse
momentum, but this investigation is left for future work.

\subsection{Impact of Activation Bitwidth}
\label{sec:act_bw_impact}

\begin{figure}
		\centering
		\begin{tikzpicture}
	\begin{semilogxaxis}[%
	xlabel=Model LUTs,
	ylabel=Accuracy \%,
	height=4cm,
	width=0.45\textwidth,
	legend pos=south east,
	scatter/classes={%
		1={mark=square*,blue},%
		2={mark=triangle*,red},%
		3={mark=o,draw=black}}]
	\addplot[scatter,only marks,%
		scatter src=explicit symbolic]%
	table[meta=label] {
x	y	label
352	83.04	1
116	83.3	1
212	83.54	1
256	83.6	1
352	83.68	1
1060	83.84	1
212	83.98	1
404	84	1
788	84.38	1
1556	84.38	1
544	84.42	1
1760	84.42	1
1892	84.44	1
412	84.52	1
13792	84.6	1
1952	84.62	1
1696	84.62	1
4452	84.7	1
7392	84.74	1
3092	84.74	1
928	84.86	1
3232	84.86	1
35264	87.42	2
11464	87.58	2
32320	88.44	2
2120	88.46	2
3520	88.56	2
4040	88.6	2
7880	88.6	2
30920	88.6	2
15560	88.7	2
9280	88.72	2
5440	88.76	2
16960	88.8	2
157760	88.88	2
133960	88.88	2
469440	88.9	2
36040	88.92	2
87240	88.92	2
36040	88.92	2
68680	88.94	2
54060	88.96	3
59840	88.98	2
59840	88.98	2
3473196	89.06	3
92480	89.06	2
89760	89.26	3
13332	89.34	3
5766816	89.36	3
51348	89.46	3
102036	89.5	3
17952	89.56	3
55968	89.58	3
106656	89.68	3
30624	89.7	3
434388	90.02	3
827796	90.14	3
721248	90.16	3
1614612	90.16	3
1114656	90.26	3
1901472	90.28	3
	};
	\legend{$\beta=1$, $\beta=2$, $\beta=3$}
	\end{semilogxaxis}
\end{tikzpicture}
        \caption{Accuracy impact of $\beta$ with model LUTs.}
		\label{fig:act-bw-effects}
\end{figure}
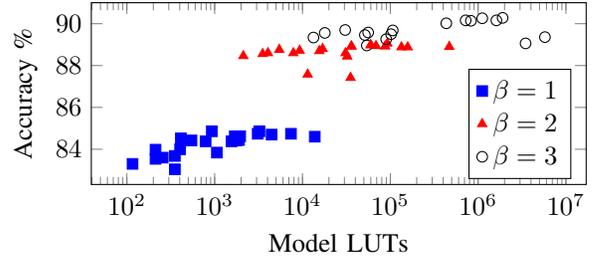

To better understand the impact of activation bitwidth ($\beta$) on accuracy,
we train a variety of topologies with different number of neurons, fan-ins ($\gamma$) and activation bitwidths ($\beta$), plotting the
accuracy on the JSC task against the analytical LUT cost in Figure~\ref{fig:act-bw-effects}, grouping
the datapoints by $\beta$.
We observe a general increase in accuracy for the larger models, and some interesting behavior regarding $\beta$.
Although binary activations generally have lower LUT cost, we observe that higher $\beta$ can yield higher accuracy at similar LUT cost.
For instance, the lowest-cost $\beta=2$ datapoint offers 88.7\% accuracy with 2120 LUTs, whereas the highest-accuracy binary
result provides 85.3\% with 7392 LUTs. A similar trend is observed for $\beta=3$, although the accuracy improvement
over $\beta=2$ is smaller.

\begin{figure}
		\centering
		\begin{tikzpicture}
	\begin{axis}[%
	xlabel=Total Neurons,
	ylabel=Accuracy \%,
	height=4cm,
	xmax=2200,
	width=0.45\textwidth,
	legend pos=south east,
	scatter/classes={%
		1={mark=square*,blue},%
		2={mark=triangle*,red},%
		3={mark=o,draw=black}}]
	\addplot[scatter,only marks,%
		scatter src=explicit symbolic]%
	table[meta=label] {
x	y	label
197	83.3	1
101	83.54	1
197	83.98	1
101	83.6	1
197	83.04	1
197	83.68	1
197	84	1
389	84.52	1
773	84.42	1
1541	84.38	1
389	84.86	1
197	83.84	1
197	84.38	1
197	84.62	1
197	84.42	1
197	84.44	1
1541	84.62	1
197	88.46	2
197	84.74	1
3077	84.86	1
773	88.56	2
3077	88.6	2
197	84.7	1
197	88.76	2
3077	84.74	1
197	88.6	2
197	88.72	2
389	87.58	2
773	89.34	3
3077	84.6	1
1541	88.7	2
773	88.8	2
389	89.56	3
1541	89.7	3
773	88.6	2
773	88.44	2
197	87.42	2
197	88.92	2
197	88.92	2
197	89.46	3
389	88.96	3
197	89.58	3
197	88.98	2
197	88.98	2
197	88.94	2
389	88.92	2
197	89.26	3
389	89.06	2
197	89.5	3
1541	89.68	3
3077	88.88	2
389	88.88	2
1541	90.02	3
3077	88.9	2
773	90.16	3
197	90.14	3
389	90.26	3
197	90.16	3
773	90.28	3
389	89.06	3
773	89.36	3
	};
	\legend{$\beta=1$, $\beta=2$, $\beta=3$}
	\end{axis}
\end{tikzpicture}
        \caption{Accuracy impact of $\beta$ with number of neurons.}
		\label{fig:n-bw-effects}
\end{figure}
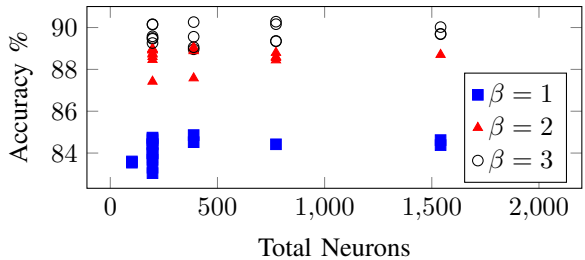

Figure~\ref{fig:n-bw-effects} presents this from another perspective, where \OurScheme{} with the same number of
neurons but different $\beta$ and $\gamma$ are clustered in columns. Training the same number of neurons with greater
activation bitwidth yields greater accuracy. Increasing the number of neurons improves the results slightly for $\beta = 1$
but brings little to no benefit for $\beta > 1$.
In general, it is difficult to estimate the effect of DNN hyperparameters on accuracy, but our experiments
indicate that the activation bitwidth $\beta$ is vital to achieving good accuracy for \OurScheme{}.

\subsection{Lossless Pruning via Synthesis Optimizations}
\label{sec:lut_reduction_synth_opts}


\begin{figure}
		\centering
		\includegraphics[width=0.5\textwidth]{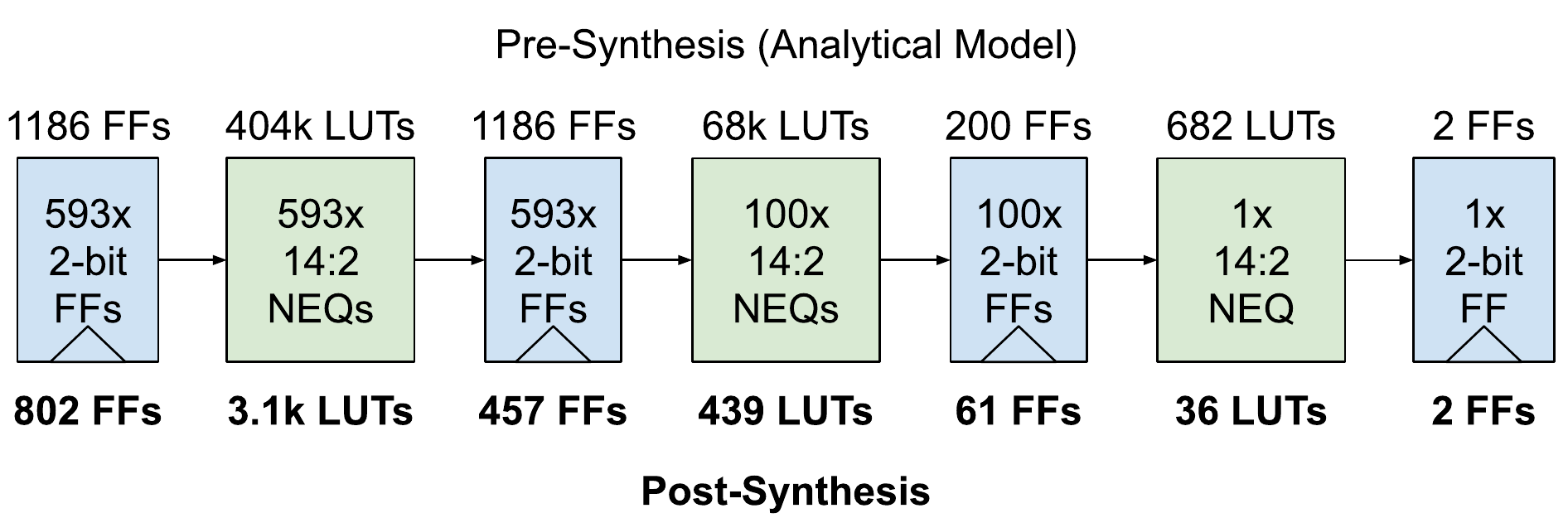}
		\caption{Post-synthesis resource breakdown for NID-S.}
		\label{fig:pre-post-synth-NID-L}
\end{figure}

There are notable differences in the LUT counts yielded by the analytical model and post-synthesis results in Table~\ref{tab:key_results},
where the post-synthesis LUT
counts are on average $39\times$ smaller than the LUT cost predicted by the analytical model.
To understand why, we take a closer look at the
post-synthesis utilization data for NID-S as reported by Vivado.
Figure~\ref{fig:pre-post-synth-NID-L} illustrates the pre- and post-synthesis resource usage per layer, and the registers between layers.
The extent of pruning is evident in both LUT and FF counts, indicating that many neurons are
removed entirely, while others are implemented with much smaller cost.
The drastic reduction in resources can be attributed to two effects.
Firstly, many NEQs have no path to output owing to a combination of fixed random
sparsity, the chosen $\gamma$ and neuron counts.
This is reflected in the post-synthesis results, which removes many unconnected neurons and registers
due to the over-provisioned topology, resulting in fewer LUTs than what is predicted by our
basic analytical model due to effective reduction of $X$ and number of neurons.
The JSC models are less over-provisioned and exhibit smaller post-synthesis savings of 1.5--8$\times$.
The second effect is the actual cost of single HBBs, which is 682 LUTs for the 14:2 HBB according to
the analytical model.
Examining the per-neuron resource costs, we observe LUT costs ranging from 10--262, as well as instances F7MUX and F8MUX primitives.
This indicates that significant savings of 2.6--68$\times$ can be achieved by applying heuristic logic
minimization on the learned HBB functions.
We leave a more in-depth study of \OurScheme{} logic minimization for future work.
%

\subsection{Limitations}
Based on our evaluation, we identify two key limitations of our current
approach.
The first is the inability to map dense networks with high fan-in neurons
commonly used in machine learning, instead requiring custom topologies that
apply fan-in restrictions via sparsity and activation quantization.
The second is the large design space with combinations of layer and neuron
counts, activation bitwidths and fan-ins that must be explored to find
suitable networks of sufficiently high accuracy and low resource cost.
Although these may be acceptable limitations for extreme-throughput applications
that commonly require specialized design effort, we hope to address these
limitations in future work to broaden the scope of \OurScheme{}.

\subsection{Comparison to Related Work}
\label{sec:comparison_to_related_work}

\begin{table}
  \caption{Comparing \OurScheme{} to related work with II=1.}
  \label{tab:relatedwork}
  \centering
  \begin{tabular}{ccccccc}
    \toprule
    Work & Accuracy & $F_{clk}$ & Latency & Resources \\
    \midrule
    JSC-L & 71.8\% & 384~MHz & 13~ns & 37.9k LUT \\
    \cite{duarte2018fast} & 75\% & 200~MHz & 50~ns & 88k LUT, 1k DSP \\
    \midrule
    NID-M & 91.30\% & 471~MHz & 10.5~ns & 15.9k LUT \\
    \cite{murovivc2019massively} & 90.1\% & 51~MHz & 19.6~ns & 51k LUT \\
    \bottomrule
  \end{tabular}
\end{table}

We compare \OurScheme{} to two prior works on extreme-throughput, fully-unfolded (II=1) FPGA implementations on these domains,
according to the metrics presented in Table~\ref{tab:relatedwork}.
For the JSC task, we take the work by Duarte~et~al.~\cite{duarte2018fast} as our baseline, which reports an accuracy of 75\% and
presents a fully-unrolled FPGA implementation of a pruned and 16-bit quantized and 30\% sparse neural network.
To focus on the implementation of the core neural network part itself, we exclude the 5-cycle softmax execution from the latency.
Our JSC-L implementation is able to offer $1.9\times$ higher throughput and $ 3.8\times$ lower latency at the cost of 3.2\% points lower
accuracy.
Since JSC-L uses less than half of the LUTs compared to \cite{duarte2018fast} and no DSP slices, the accuracy could be potentially
increased further by using a larger \OurScheme{}-style network and more resources.
For the NID task, we compare against Murovic~et~al.~\cite{murovivc2019massively}, who implement a dense binarized neural network achieving 90.1\% accuracy
as a fully-unfolded, combinatorial circuit without any registers.
The \OurScheme{} NID-M implementation outperforms their solution in terms of all metrics, offering 1.2\% points higher accuracy, $9.2\times$
higher throughput and $ 1.9 \times$ lower latency using $ 3.2\times $ fewer LUTs.


\section{Conclusion and Future Work}

In this work, we have investigated how DNN topologies that map well to
FPGA building blocks can be constructed to obtain efficient implementations
for extreme-throughput applications.
Noting that quantized neurons with limited fan-in can be converted into small
truth tables, we have proposed a flow to design sparse, quantized topologies
that map to highly efficient FPGA implementations.
On two tasks with extreme-throughput requirements, we were able to demonstrate
implementations with competitive accuracy, low latency and throughput in the
hundreds of millions of inferences per second.

\OurScheme{} opens up a wide array of possible future work owing to its
cross-stack nature which spans machine learning, compilation and FPGA synthesis.
On the machine learning side, we plan to explore training-time methods to
increase the accuracy of sparse and quantized topologies, exploring the
possibilities for sparse convolutions, as well as mixing \OurScheme{}-style
layers with more conventional ones in the same topology in order to increase the
accuracy and apply it to more difficult problems.
On the tooling and hardware synthesis side, we hope to further study the benefits of
heuristic logic minimization to design more accurate analytical cost models, as well
as techniques to synthesize larger \OurScheme{} models quickly and efficiently.

\bibliographystyle{IEEEtran}
\bibliography{IEEEabrv,refs.bib}


\end{document}